\documentclass[12pt,a4paper,palatino]{article}
\usepackage{cite}
\usepackage{graphicx}
\usepackage{amssymb}
\usepackage{epsfig}
\usepackage[tmargin=1cm,bmargin=1cm,lmargin=2cm,rmargin=2cm]{geometry}

\renewcommand{\thefootnote}{\arabic{footnote}}
\newcommand{\sm}{\rm SM}
\newcommand{\bz}{\cal{BZ}}
\newcommand{\uu}{\cal{U}}
\nopagebreak

\begin{document}

\pagestyle{empty}
\baselineskip 22pt
\vspace*{-1in}
\renewcommand{\thefootnote}{\fnsymbol{footnote}}
\begin{flushright}
SINP/TNP/2007/29\\
{\tt arXiv:0709.2478 [hep-ph]}\\
{\tt published in Phys.Lett.B657:198-206,2007.}
\end{flushright}
\vskip 65pt
\begin{center}
{\Large \bf Unparticle Physics in di-photon production at the LHC} 

\vspace{8mm}
{\bf
M. C. Kumar$^{a,b}$
\footnote{mc.kumar@saha.ac.in}, 
Prakash Mathews$^a$
\footnote{prakash.mathews@saha.ac.in},
V. Ravindran$^c$
\footnote{ravindra@mri.ernet.in},
Anurag Tripathi$^c$
\footnote{anurag@mri.ernet.in}
}\\
\end{center}
\vspace{10pt}
\begin{flushleft}
{\it
a) 
Saha Institute of Nuclear Physics, 1/AF Bidhan Nagar,
Kolkata 700 064, India.\\

b)
School of Physics, University of Hyderabad, Hyderabad 500 046, India.\\

c)
Regional Centre for Accelerator-based Particle Physics,\\ 
~~~Harish-Chandra Research Institute,
 Chhatnag Road, Jhunsi, Allahabad, India.\\
}
\end{flushleft}
\vspace{10pt}
\begin{center}
{\bf ABSTRACT}
\end{center}

We have considered the diphoton production in the unparticle physics at the LHC.  The 
contributions of spin-0 and spin-2 unparticles to the di-photon production 
are studied in the invariant mass and other kinematical distributions, 
along with their dependencies on the model dependent parameters.  It is found that the 
signal corresponding to the unparticles is significant for moderate values of
the couplings.

\vskip12pt
\vfill
\clearpage

\setcounter{page}{1}
\pagestyle{plain}

\section{Introduction}
Banks and Zaks \cite{bz} studied gauge theories with non-integral number,
$N_{F}$, of Dirac fermions, such that the two loop beta function vanishes.
At this nontrivial infra-red (IR) fixed point, the theory is scale invariant and does not have a
particle interpretation.  Motivated by Banks and Zaks, 
Georgi \cite{hep-ph/0703260} proposed the following scheme:
Theory at very high energy contains the fields of the standard model (SM) and
fields of a sector called Banks-Zaks $\bz$ sector, with a nontrivial IR
fixed point. These two sectors interact through exchange of particles with a
large mass scale $M_{\cal U}$.   Below $M_{\cal U}$ the couplings have generic
form
\begin{eqnarray}
\frac{1}{M^k_{\cal U}} O_{\sm} O_{\bz} ~,
\label{eq1}
\end{eqnarray}
where $O_{\sm}$ and $O_{\bz}$ are operators built out of the standard model and
the $\bz$ fields respectively. Scale invariance in the $\bz$ sector emerges at 
energy scale $\Lambda_{\cal U}$. In the effective theory below
$\Lambda_{\cal U}$ the interaction of (1) matches onto
\begin{eqnarray}
C_{\cal U} \frac{\Lambda_{\cal U}^{d_{\bz} - d_{\cal U}}}{M^k_{\cal U}} O_{\sm}
O_{\cal U} ~,
\label{eq2}
\end{eqnarray}
where $d_{\cal U}$ is the scaling dimension of unparticle operator $O_{\cal U}$.
%
$M_{\cal U}$ should be large enough that its coupling to $\sm$
be sufficiently weak. Few of the generic operators that can describe the interaction of unparticle fields 
with those of the SM are found to be
\begin{eqnarray}
\frac{\lambda_s}{\Lambda^{d_{\cal U}}_{\cal U}} T_{\mu}^\mu ~O_{\cal U} ,
\qquad
\frac{\lambda_v}{\Lambda^{d_{\cal U}-1}_{\cal U}} \bar{\psi} \gamma_\mu \psi
~O^\mu_{\cal U} ,
\qquad
\frac{\lambda_t}{\Lambda^{d_{\cal U}}_{\cal U}} T_{\mu\nu} ~O^{\mu\nu}
_{\cal U} ~.
\label{eq3}
\end{eqnarray}
The dimensionless coupling $\lambda_{\kappa}$ corresponds to the
unparticle operator $O^\kappa_{\cal U}$, where $\kappa=s,v,t$
refers to the scalar, vector and tensor operators respectively.
$T_{\mu\nu}$ is the energy momentum tensor of the $\sm$.
These operators are Hermitian and 
transverse.

The unparticle propagator \cite{hg2-kingman} is given by
\begin{eqnarray}
\int d^4 x e^{i P x} <0|T O^{\kappa}_{\cal U} (x) O^{\kappa}_{\cal U} (0)|0>
&=&\frac{i A_{d_{\cal U}}}{2 \sin(d_{\cal U} \pi)} \frac{B_{\kappa}}
{(-P^2-i \epsilon)^{2-d_{\cal U}}} ~,
\label{eq4}
\end{eqnarray}
where, $B_{\kappa}$ depends on the Lorentz structure of the operator
$O_{\cal U}$ as given below:
\vskip 0.1cm
\begin{center}
\begin{tabular}{ccc}
$O_{\cal U}$ & \qquad \qquad &1\\
{}&{}\\
$O^\rho_{\cal U}$ & & $\eta_{\mu\nu}(P)= -g_{\mu\nu}+ \frac{P_\mu P_\nu}{P^2}$
\\
{}&{}\\
$O^{\rho \sigma}_{\cal U}$ & & $B_{\mu\nu\alpha\beta}(P)=\frac{1}{2} \left(
\eta_{\mu\alpha} 
\eta_{\nu\beta}+\eta_{\mu\beta} \eta_{\nu\alpha}- 
\frac{2}{3}\eta_{\mu\nu} \eta_{\alpha\beta} \right)$ ~~.
\end{tabular}
\end{center}
\vskip 0.1cm
The constant $A_{d_{\cal U}}$ is given by
\begin{eqnarray}
A_{d_{\cal U}}=\frac{16 \pi^{5/2}}{(2 \pi)^{2 d_{\cal U}}}
\frac{\Gamma(d_{\cal U}+1/2)}{\Gamma(d_{\cal U}-1)\Gamma(2 d_{\cal U})}~~.
\end{eqnarray}
where $1 < d_{\cal U} < 2$.

If $\Lambda_{\cal U}$ is of order TeV, unparticle dynamics can be 
seen at the Large Hadron Collider (LHC) through various high energy scattering 
processes and hence the phenomenology with it will be interesting.
Several detailed studies on the phenomenology of unparticle physics have been reported
in the recent past exploring the possibility of explaining the known experimental
results and also constraining the parameters of the model.
Supersymmetric scenarios were taken up in \cite{susy} and effects on 
cosmology and astrophysics have been considered in \cite{hooman}.  For studies in 
flavor physics and CP violations, see \cite{flavor}. In the context of neutrino physics,
the unparticle physics has been studied by authors of \cite{neutrino}.  Study of \cite{Stephanov} 
showed that unparticles can be represented as an infinite tower of 
massive particles with controllable mass--squared spacing $\Delta^{2}$,
and that pure unparticles cannot decay, while for small $\Delta$ the 
decay is possible.  In \cite{nath}, lowest order 
ungravity correction to the Newtonian gravitational potential has been computed and 
it is found that $ 1<d_{U}<2 $ leads to modification of the inverse square law 
with $r$ dependence in the range $1/r^2$ and $1/r^4$ and also explored on how to 
discriminate extra dimension models and ungravity models  
in sub milli-meter tests of gravity.  It was found in \cite{kikuchi} that the
unparticles can modify the coupling between Higgs and a pair of gluons/photons and
its effects can be observed in di-photon productions through Higgs decay processes
at the LHC. 
Unparticle contributions to mono-jet, 
and di-photon production at $e^{+}e^{-}$ colliders are now known \cite {cheung}.  

Drell-Yan production at hadron collider via unparticles has been reported in \cite{PM},
where we had restricted ourselves to scalar and tensor unparticles to find out
the plausible region of the parameter space to see their effect.  We had also incorporated
next to leading order QCD effects to stabilise our results against both
higher order corrections and scale variations.   
We list in \cite{others} most of the articles in the context of unparticle physics.
In this paper, we have studied the impact of unparticle fields on one of the important
processes, namely di-photon production.

\section{The Diphoton production}

The production of di-photon system is one of the important processes at the hadron
colliders and has been used to do precision study of the Standard Model (SM).  Also 
it provides a laboratory for probing new physics.  In the SM, this process has been 
studied in great detail including higher order QCD 
\cite{Binoth:1999qq} effects.  
The soft gluon effects through threshold resummation have also been incorporated (see \cite{yuan}).   
In the context of physics beyond the SM, this process has played 
an important role in constraining parameters of various models.  For example, models with 
large extra dimensions can be probed using the di-photon signals \cite{eboli}    
 (See also \cite{ex} for the bounds coming from the Tevatron).
In this paper,  we study the effect of unparticles on
various kinematical distributions of di-photon system produced at hadron colliders. 
At hadron colliders, the di-photon system can be produced through 
\begin{eqnarray}
P_1(p_1) + P_2(p_2) \to \gamma(p_3) + \gamma(p_4) + X(p_X) ~,
\end{eqnarray}
where $P_i$ are the incoming hadrons with momenta $p_i$ and $X$ is the final inclusive hadronic state.  
The hadronic cross sections can be obtained by convoluting
the partonic cross sections $d \hat \sigma^{ab}$ with the the appropriate incoming 
parton distribution functions $f^P_a$:
\begin{eqnarray}
d\sigma(P_1 P_2 \to \gamma \gamma X) = \sum_{a,b=q,\bar{q},g}
\int f^{P_1}_a (x_1) f^{P_2}_b (x_2)
d\hat\sigma^{ab} (x_1 , x_2 ) dx_1 dx_2 ~.
\end{eqnarray}
Here, $x_1$ and $x_2$ are the momentum fractions of the incoming partons in the hadrons
$P_1$ and $P_2$ respectively.

The unparticle model has spin-0 and spin-2 unparticles.  Thus, 
di-photon can be produced 
through $q \bar q$ annihilation as well as the $gg$ fusion subprocesses with scalar and
tensor unparticles appearing as propagators (see Eq.(\ref{eq4})).   
These partonic sub-processes occur
at leading order in couplings $\lambda_s$ and $\lambda_t$.  
At this order spin-1 unparticle does not contribute.

The matrix element squared of the partonic subprocess due to
scalar unparticle is found to be
\begin{eqnarray}
\nonumber
| \overline{M}_{q \bar{q}}|^2&=&\frac{1}{8N_c} ~\lambda_s^4  ~\chi_{\cal U}^2
\left( \frac{s}{\Lambda_{\cal U}^2}\right)^{2d_{\cal U} -1} \, ,\\
| \overline{M}_{gg}|^2& = & \frac{1}{8(N_c^2-1)} \frac{1}{4}  ~\lambda_s^4 
 ~\chi_{\cal U}^2
\left(\frac{s}{\Lambda_{\cal U}^2}\right)^{2 d_{\cal U}} \, .
\label{msqs}
\end{eqnarray}
Similarly, we have for tensor unparticle exchange:
\begin{eqnarray}
\nonumber
| \overline{M}_{q \bar{q}}|^2&=& \frac{1}{8N_c} \Bigg[ e^4 Q_f^4 8
\left(\frac{u}{t} +\frac{t}{u} \right)\\ \nonumber
{}   && -8e^2Q_f^2 ~\lambda_t^2 ~\chi_{\cal U} ~\cos(d_{\cal U} \pi)
\left(\frac{s} {\Lambda^2_{\cal U}} \right)^{d_{\cal U}} \frac{1}{s^2}
(u^2+t^2) \\ \nonumber
{}   && +2 ~\lambda_t^4  ~\chi_{\cal U}^2 \left(\frac{s}
{\Lambda_{\cal U}^2}\right)^{2d_{\cal U}}
\frac{1}{s^4} t u (u^2+t^2) \Bigg]  ~,\\
|\overline{M}_{gg}|^2&=&\frac{1}{8(N_c^2-1)}2  ~\lambda_t^4  ~\chi_{\cal U}^2
 ~\left(\frac{s}{\Lambda_{\cal U}^2}\right)^{2 d_{\cal U}} \frac{1}{s^4}
(u^4+t^4) ~.
\label{msqt}
\end{eqnarray}
where, $Q_f$ is the electric charge of the parton of flavour $f$, 
$\chi_{\cal U}=A_{d_{\cal U}}/(2 \sin (d_{\cal U} \pi) )$ and $N_c$
is the number of colors.   The variables $s,t$ and $u$ are the standard partonic
Mandelstam invariants.  Notice that only tensor unparticles interfere
with the SM subprocess amplitudes.  
In the above matrix elements (Eqs. (\ref{msqs},\ref{msqt})), 
we have already done spin and colour averages and included the 
correct symmetry factor coming from the identical nature of the final state photons.
It may be noted here that the spin-0 and spin-2 unparticles do not interfere.

As was pointed out in \cite{bern}, the $gg \to \gamma \gamma$ box 
contribution at order $\alpha_s^2$ could give a sizeable effect
at the LHC in comparison to other contributions to this order due to large
gluon flux.  There are numerous diagrams that contribute to
order $\alpha_s^2$, but the authors of \cite{bern} have argued 
that the dominant 
contribution comes from the box diagram.  For unparticle searches 
it is plausible that the interference of the $gg$ unparticle contributions 
with the SM diagrams at order $\alpha_s^2$ could be sizeable.  
In the context of large extra-dimensional models, these interference 
contributions enhance the effect at small $Q$ regions \cite{eboli}.  
For our present analysis we have restricted to only the LO in QCD.  

\vspace{.3cm}
\begin{figure}[htb]
\centerline{
\epsfig{file=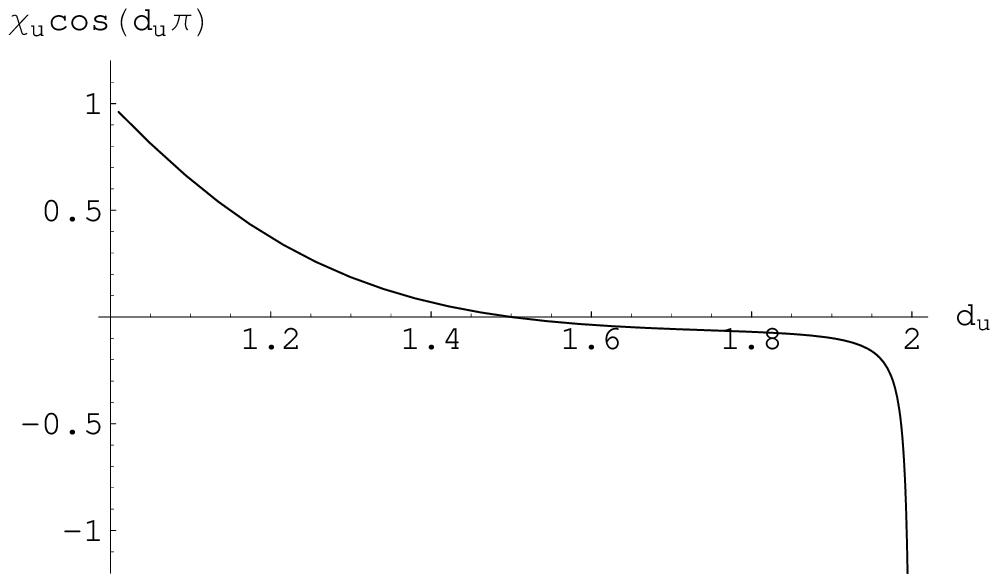,width=8cm,height=5cm,angle=0}
\epsfig{file=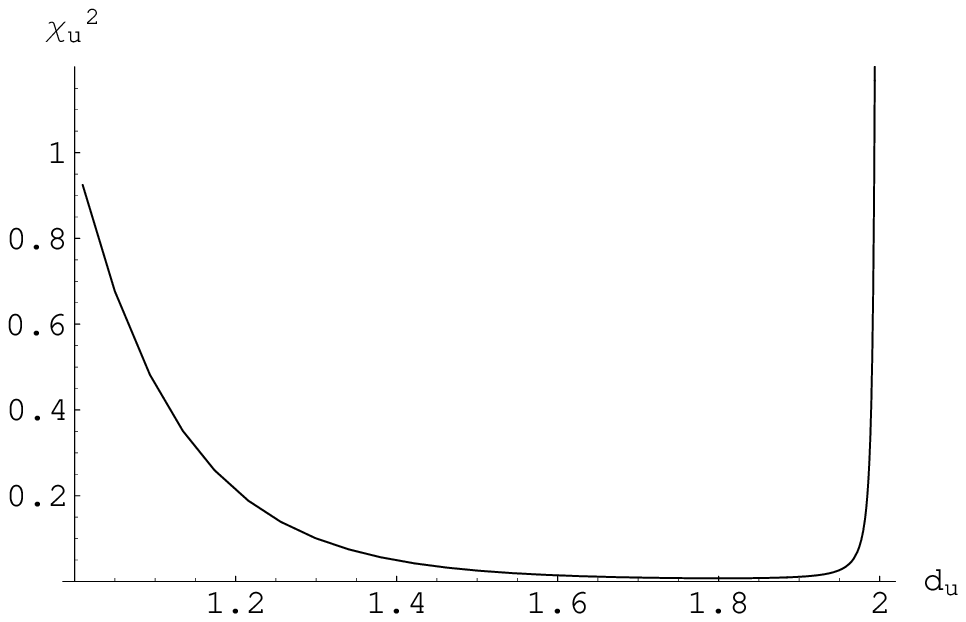,width=8cm,height=5cm,angle=0}
}
\vspace{.1cm}
\caption{The function $\chi_{\uu}~ \cos (\pi d_{\uu})$ (left) and
$\chi_{\cal U}^2$ (right) showing its variation with the scaling
dimension $d_{\cal U}$ of the unparticle operator.}
\label{chiu}
\end{figure}

Unitarity imposes 
constraint \cite{mack} on the conformal dimension 
\footnote{There exist no known examples of scale invariant local field 
theories that are not conformally invariant (Y.~Nakayama in \cite{susy}).} 
of these operators, which for scalar unparticle is $d_{\uu}>1$.  
Though this constraint restricts the scalar unparticle sector,
for our numerical analysis we have considered $d_{\uu}>1$ as the general constraint for 
other unparticle operators (say tensor unparticle) as well.  
Before, we present the effects of unparticles on various distributions
of di-photon system at the LHC, we discuss the coefficients
$\chi_{\cal U} \cos (d_{\cal U} \pi)$ and  
$\chi_{\cal U}^2$ that enter the interference and direct unparticle contributions 
respectively.  In Fig.~(\ref{chiu}), we have plotted them
against the scaling dimension $d_{\uu}$ of the unparticle operator.  
$\chi_{\uu}$ is negative when $1<d_{\uu}<2$ and  
singular as $d_{\uu} \to 2$.  As $d_{\uu} \to 1$, 
$\chi_{\uu}$ approaches a limiting value,  here both $\chi_{\uu}^2$ and 
$\chi_{\uu} \cos (d_{\uu} \pi)$ are positive and large and as we go
below $d_{\cal U}=1.01$, the variation is found to be mild.
In the plateau region, where $1.3 <d_{\cal U}<1.9$, these functions are 
almost constant and relatively small.  
We avoid the region $1.9 \le d_{\cal U} \le 2.0$ where $\chi_{\cal U}$ is
very rapidly increasing.
Hence, in this region, the unparticle effects can not be probed.
With this information, the value of $d_{\cal U}$ is chosen 
in such a way that the unparticle effects can been seen at the LHC energy.  

The couplings of the unparticle operators to the SM fields are given by
\begin{eqnarray}
\lambda_\kappa=C^\kappa_{\uu} \left(\Lambda_{\uu} \over
M_{\uu} \right)^{d_{\bz}} {1 \over M_{\uu}^{d_{\sm}-4}} \, ,
\end{eqnarray}
A priori we have no information on any of the parameters in the
above equation.  For our numerical analysis we have taken $\lambda_\kappa$ in the
range $0.4 \le \lambda_\kappa < 1$, so that the unparticle effects are treated as perturbation.
The other parameter that appears in this model
is $\Lambda_{\cal U}$ which we choose to be $1$ TeV.

In the following, we will study the effects of scalar and tensor unparticles separately.
We will analyse these effects only for the LHC with $\sqrt{S}=14$ TeV.  A similar analysis for the Tevatron 
can be done with our numerical code that incorporates all the analytical results presented in
this paper.  We have considered four different distributions of the photons
in the final state to unravel the effects coming from the unparticles.  They are 
(a) invariant mass distribution $d\sigma/dQ$, where $Q$ is the invariant mass of the di-photon system,
(b) angular distribution $d\sigma/d \cos ~\theta^*$, 
(c) the rapidity ($Y$) distributions of 
the di-photon system and (d) rapidity ($y^\gamma$) distributions of 
the individual photons. 
We have imposed the cuts: 
rapidity $|y^{\gamma}| < 2.5$, and transverse momentum of the photons
$p^\gamma_T > 40$ GeV \cite{tdr} for all the distributions that we have reported here in order to
make our predictions for an environment which is as close as possible to that of the experiment.  
Moreover, for the invariant mass distribution,
in order to suppress the SM background and also to enhance the signal 
we have imposed an angular cut on the photons
$|\cos ~\theta_{\gamma}| < 0.8$, where $\theta_{\gamma}$ is the 
angle of the photons in the lab frame.  
Similarly, for the angular and rapidity distributions, to suppress the background, we have
considered only those events that satisfy the constraint $Q > 600$ GeV.
For all our plots, we have used MRST 2001 leading order (LO) parton density sets \cite{Martin:2002dr}.

\subsection{Invariant mass distribution}

\begin{figure}[htb]
\centerline{
\epsfig{file=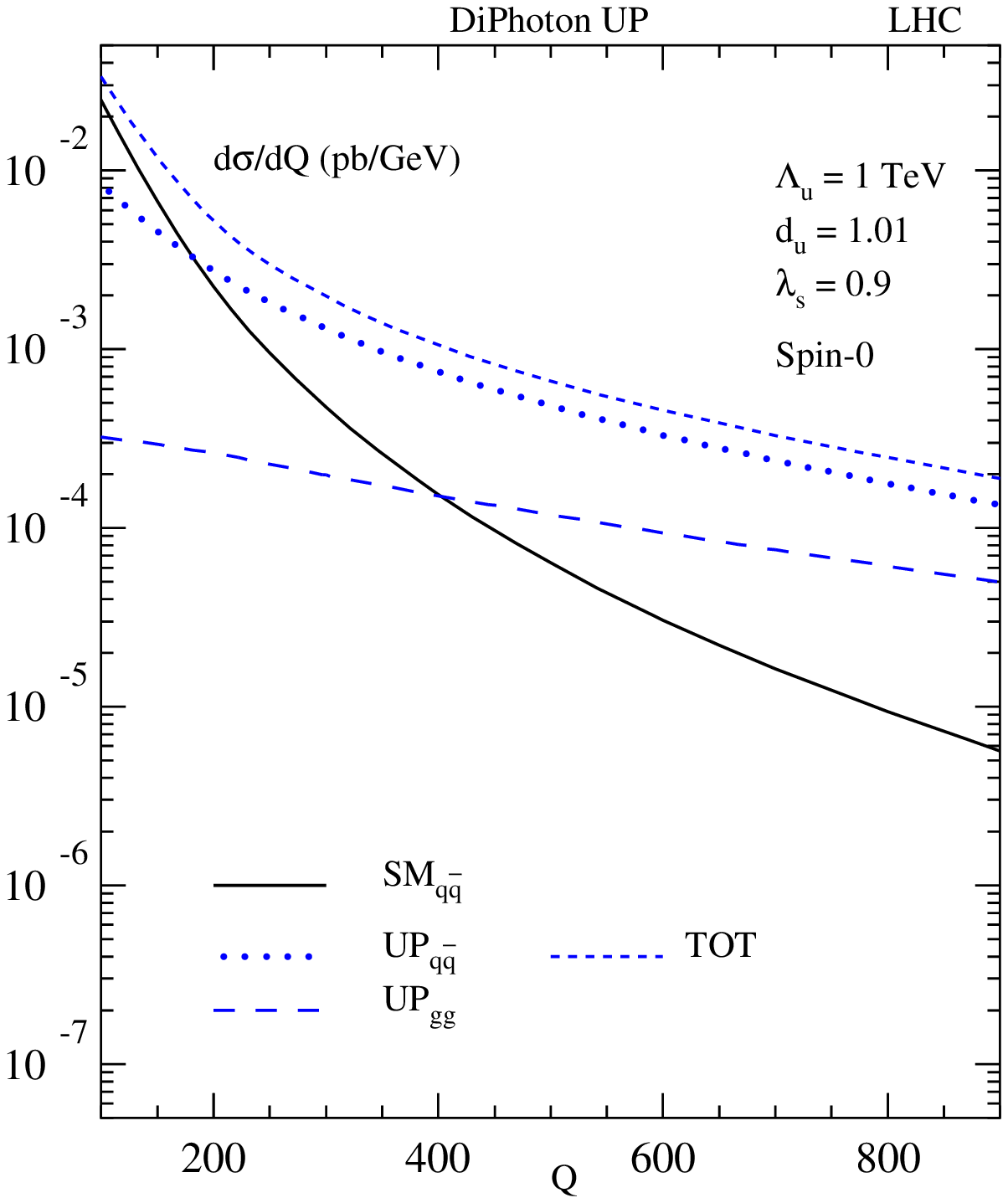,width=8cm,height=9cm,angle=0}
\epsfig{file=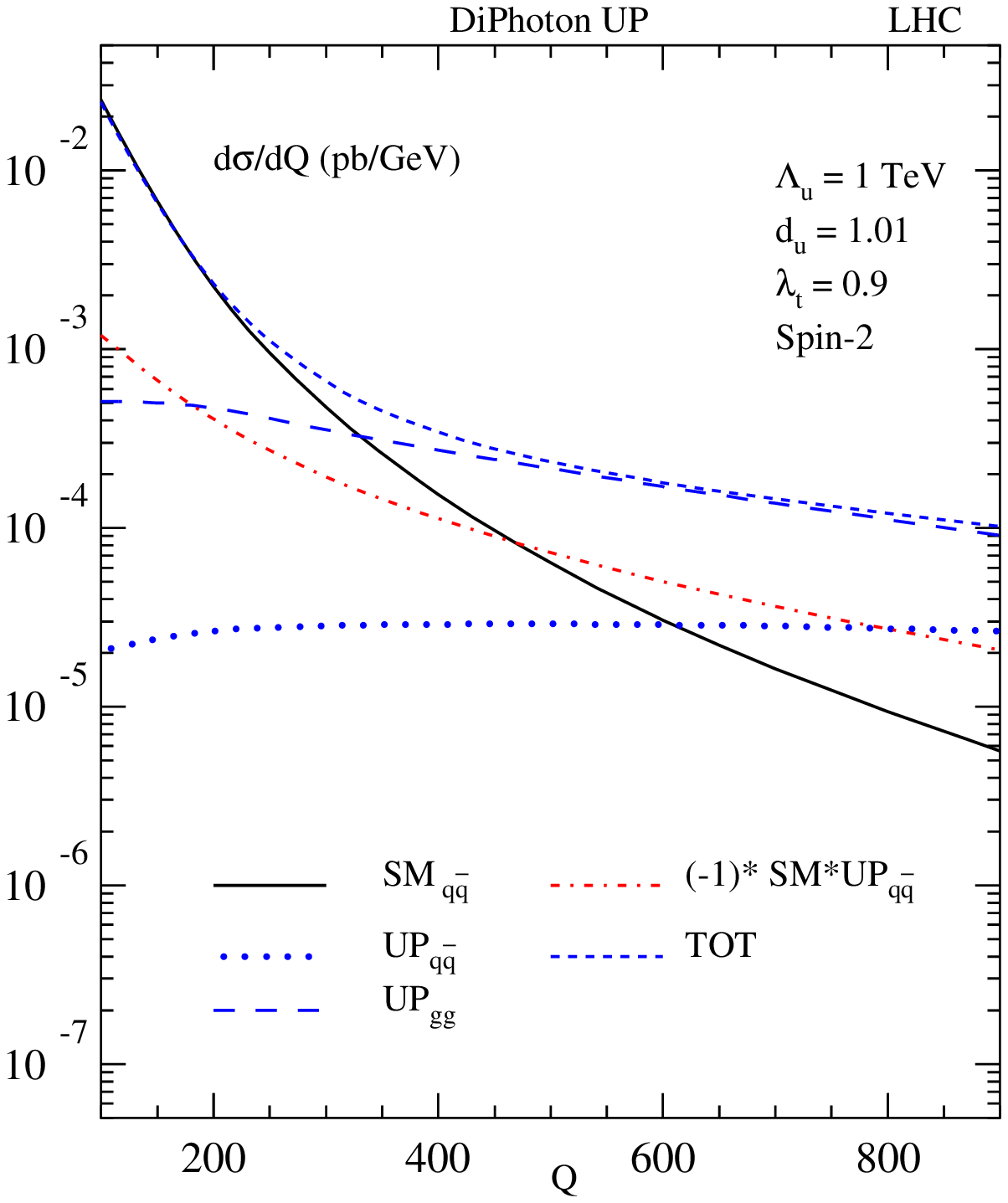,width=8cm,height=9cm,angle=0}}
\caption{ {The contribution of the various sub processes to the di-photon
production in the invariant mass distribution via scalar (left) and tensor
 (right) $s$-channel processes, with $d_{\cal U}=1.01$ and
$\Lambda_{\cal U}=1$ TeV. The scalar and tensor couplings are taken to be
$\lambda_{s,t}=0.9$.  We imposed an angular cut
$|\cos ~\theta_\gamma| < 0.8$ on the photons to suppress the SM background.}}
\label{sub}
\end{figure}

In this section, the invariant mass distribution $d\sigma/dQ$ is studied, where 
$Q^2=(p_3 +p _4 )^2$. 
In Fig.\ref{sub} we have plotted this distribution 
including the effects of scalar (left panel) and tensor (right panel) unparticles for 
$Q$ between $100 < Q < 900$ GeV.
Here we have chosen $d_{\cal U} = 1.01$ and $\Lambda_{\cal U} = 1$ TeV.  
With this choice of parameters, we find that the unparticle effects can be seen 
only in the large $Q$ region.  
In addition, we have presented different contributions coming from 
various sub-processes to the cross section for both spin-0 and spin-2 unparticles to
see their effects separately.  In the spin-0 case, the quark anti-quark initiated process
dominates over the gluon initiated process due to higher power of scale in the later case.
On the other hand, we see the opposite behavior for the spin-2 case.  In the spin-2
case, this behavior can be understood 
by noticing that the gluon fluxes are much larger compared to quark anti-quark fluxes 
at the LHC energies  
even though the couplings ($\lambda_\kappa$) for both quark anti-quark (pure unparticle contribution) 
and gluon initiated processes are same.  
In addition, the interference term, being negative reduces the contribution coming from 
quark anti-quark channel.  Notice that there is no such contribution from the spin-0 unparticle. 

\begin{figure}[htb]
\centerline{
\epsfig{file=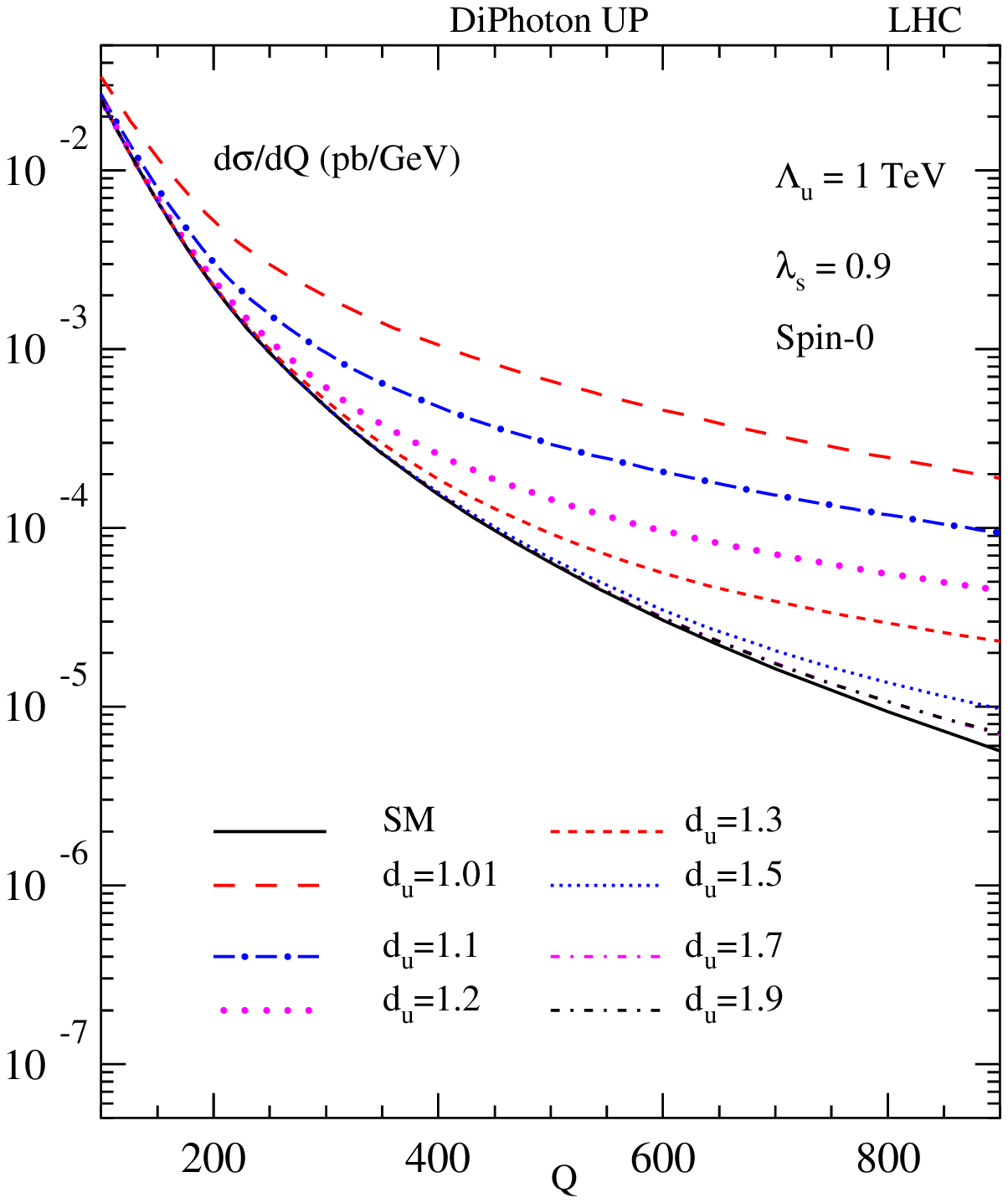,width=8cm,height=9cm,angle=0}
\epsfig{file=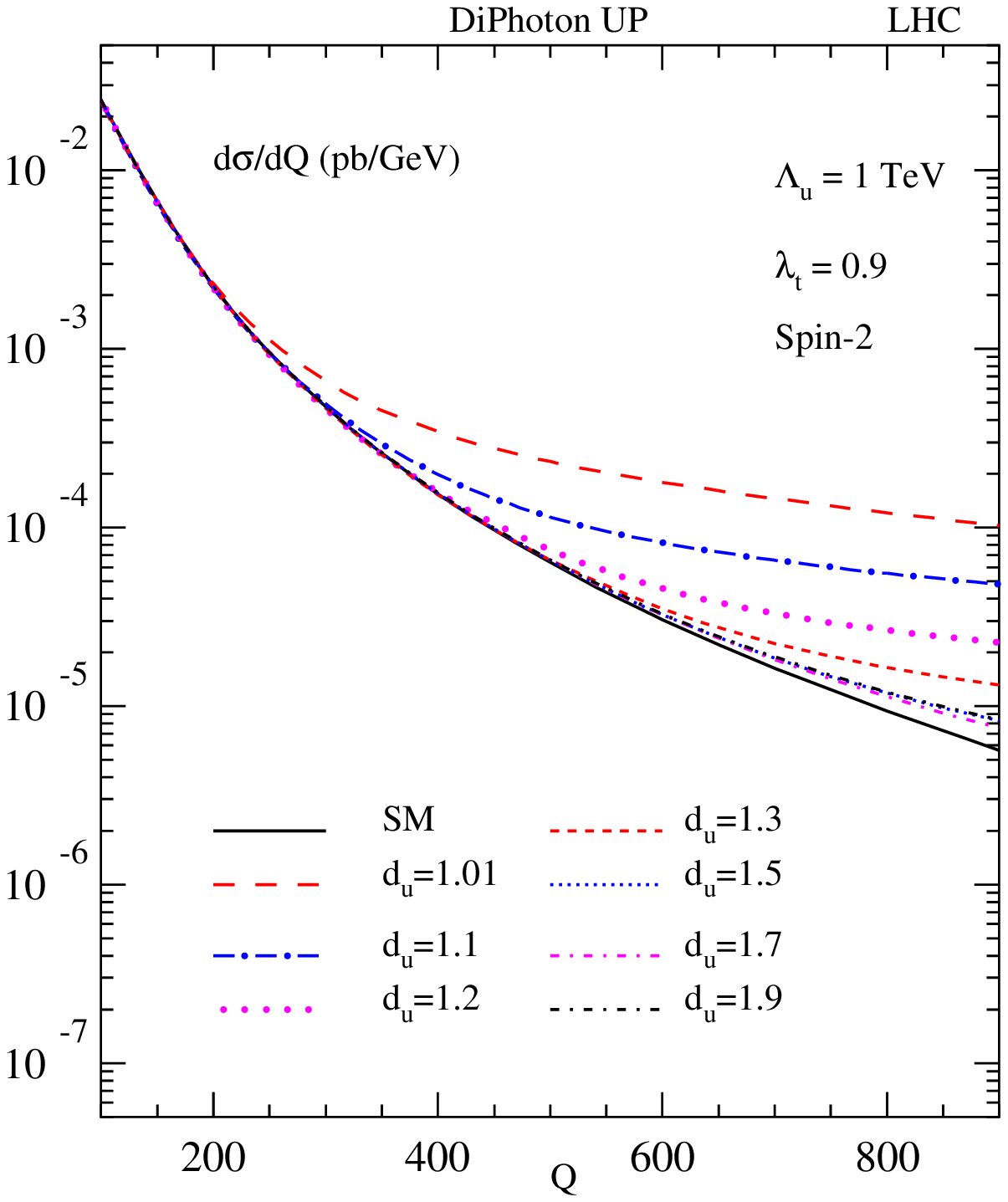,width=8cm,height=9cm,angle=0}}
\caption{ {Invariant mass distribution plotted for different values of
$d_{\cal U}$ for spin-0 (left) and spin-2 (right) with
$\Lambda_{\cal U}=1$ TeV and $\lambda_s,\lambda_t=0.9$,
with an angular cut on the photons $|\cos ~\theta_\gamma| <0.8$.}}
\label{duvariation}
\end{figure}

In Fig.\ref{duvariation}, we show the variation of the invariant mass distribution with 
respect to the scaling dimension $d_{\uu}$ of the scalar
and tensor unparticle operators, for $\Lambda_{\uu} = 1$ TeV.  
As expected, we find that the unparticle effects show up significantly when  
the value of $d_{\cal U}$ decreases. 
When $d_{\cal U}$ is around $1.9$, the unparticle effects are completely washed away.
The interference term in the spin-2 unparticle case 
gives large negative contribution only in the region where 
$d_{\uu}$ is less than $1.3$ (see Fig.~\ref{chiu}). 

\begin{figure}[htb]
\centerline{
\epsfig{file=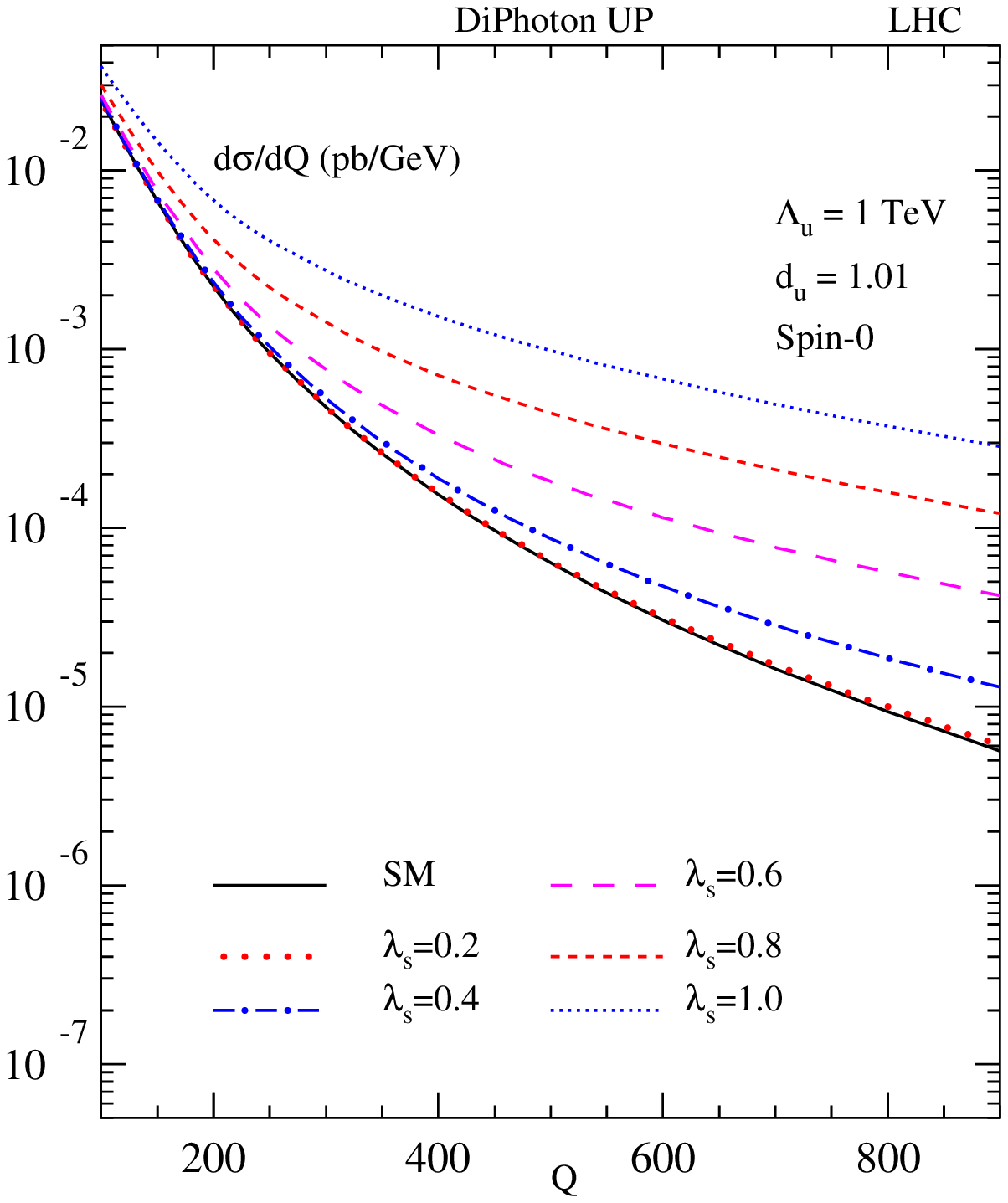,width=8cm,height=9cm,angle=0}
\epsfig{file=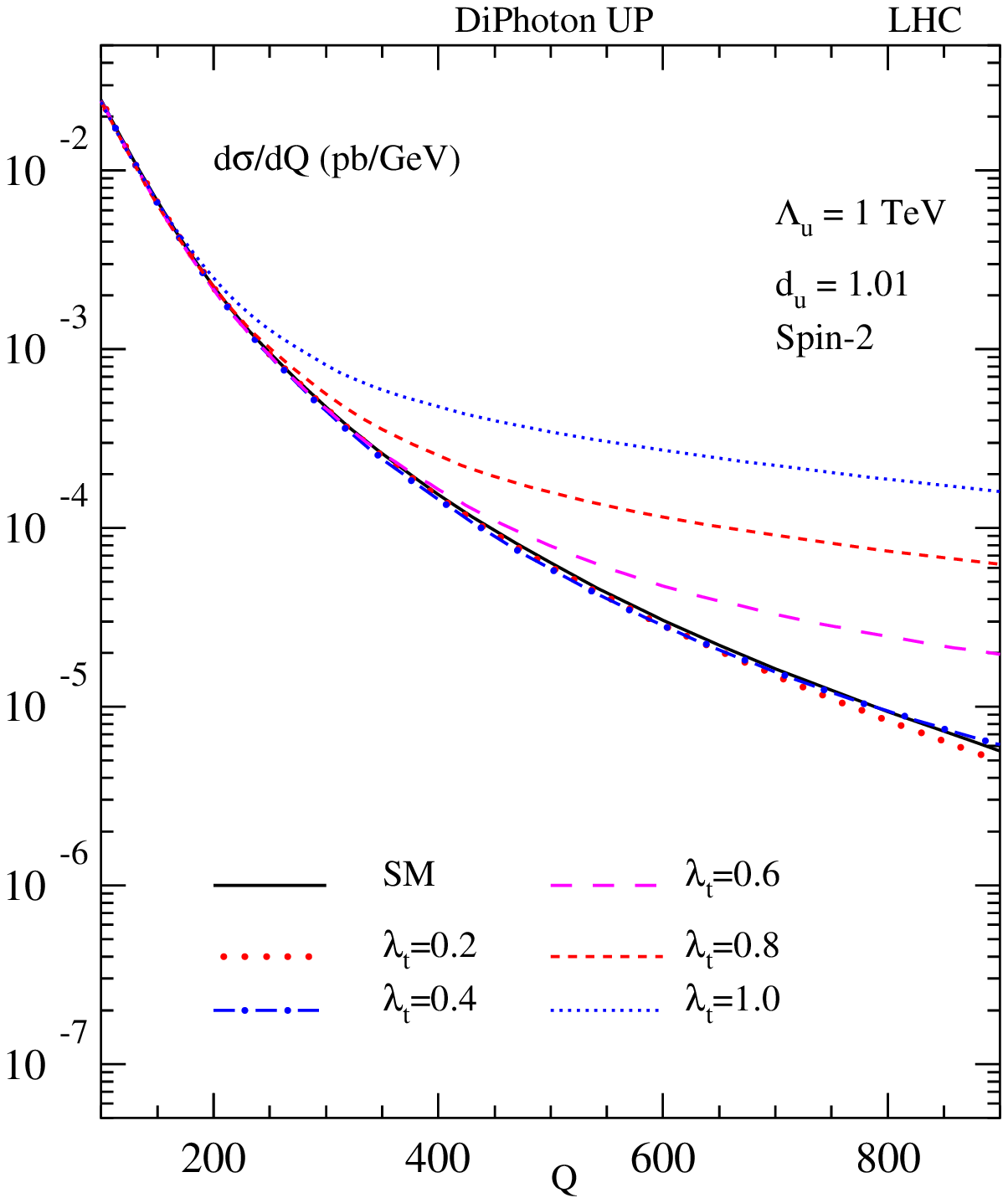,width=8cm,height=9cm,angle=0}}
\caption{ {Invariant mass distribution is plotted for various values of
the coupling $\lambda_s$ and $\lambda_t$  for spin-0 (left) and spin-2
(right) respectively with $\Lambda_{\cal U}=1$ TeV and $d_{\cal U}=1.01$,
with an angular cut on the photons $|{\cos}~\theta_{\gamma}| <0.8$.}}
\label{lambda}
\end{figure}

We have also studied the effects of $\lambda_t$ and $\lambda_s$ variations on the distributions.
They are shown in Fig.~{\ref{lambda}} for $d_{\cal U}=1.01$. 
From Fig.~{\ref{duvariation}} and Fig.~{\ref{lambda}}, we see that  
in the region where $d_{\cal U}$ is below $1.1$ and $\lambda_s$ above  $0.6$
the scalar unparticle contribution is substantial even at low energies. 

\begin{figure}[htb]
\centerline{
\epsfig{file=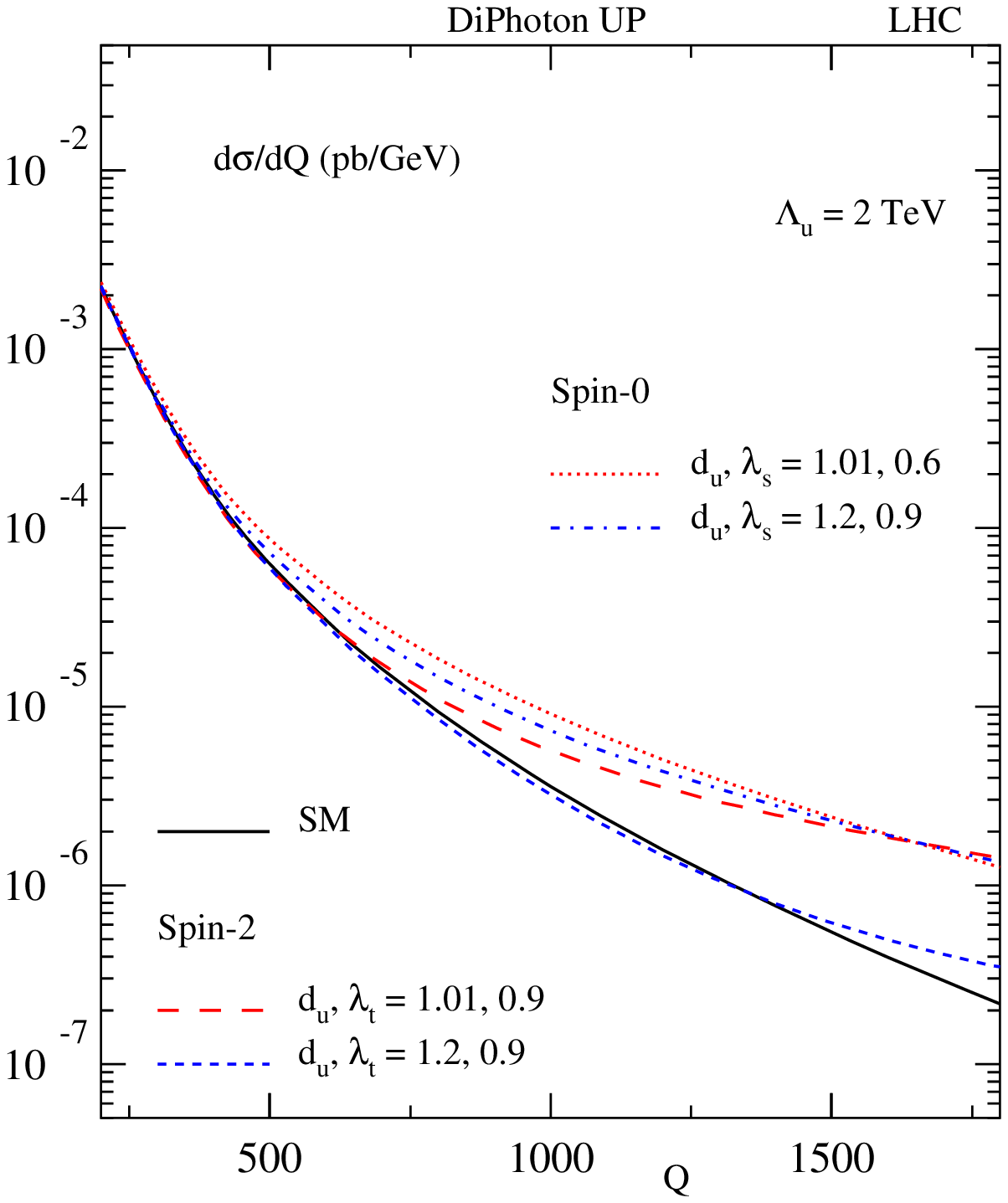,width=8cm,height=9cm,angle=0}}
\caption{ {Invariant mass distribution $d\sigma/dQ$ for the di-photon
production with $\lambda_s=0.6$ and $\lambda_t=0.9$ for $\Lambda_{\cal U}=2$
TeV for both $d_{\cal U}=1.01$  and  1.2. We have imposed an angular
cut $| \cos ~\theta_{\gamma}| <0.8$ on the photons.}}
\label{Lambdavar}
\end{figure}

The invariant mass distribution for a higher value of the scale $\Lambda_{\uu} =2$ TeV is
plotted in Fig.~\ref{Lambdavar} for various values of $\lambda_{\kappa}$ and $d_{\uu}$. 
Due to the factor $\Lambda_{\uu}^{-d_{\uu}}$ in Eqs.(\ref{msqs},\ref{msqt}) 
the cross sections are suppressed as we increase $\Lambda_{\uu}$.
In the rest of the paper, we choose $\lambda_s = 0.6$, 
$\lambda_t = 0.9$, $d_{\cal U}=1.01$ for $\Lambda_{\cal U} =1$ TeV to study other distributions.

%
%

\subsection{Angular distribution}

      The angular distribution $d\sigma/d{\cos}~\theta^*$ is
studied in the center of mass frame of the final state photons.  The angle
$\theta^*$ is defined by
\begin{eqnarray}
{\cos}~\theta^* = \frac{p_1. (p_3 - p_4)}{p_1. (p_3 + p_4)} \, ,
\end{eqnarray}
where $p_3$ and $p_4$ are the final state photon momenta.  
The distributions are plotted in the range $ -0.95 < {\cos}~\theta^* < 0.95$
for both spin-2 and spin-0 cases. 
We take $d_{\cal U}=1.01$, $\Lambda_{\cal U}=1$ TeV, 
$\lambda_t = 0.9, 0.4 $ and $\lambda_s = 0.6,0.4$. 
This angular distribution is computed in the region where $Q$ between 
$600$ GeV and $0.9 \Lambda_{\uu}$ contributes.  In this region, the unparticle 
effects are expected to be large.
\begin{figure}[htb]
\centerline{
\epsfig{file=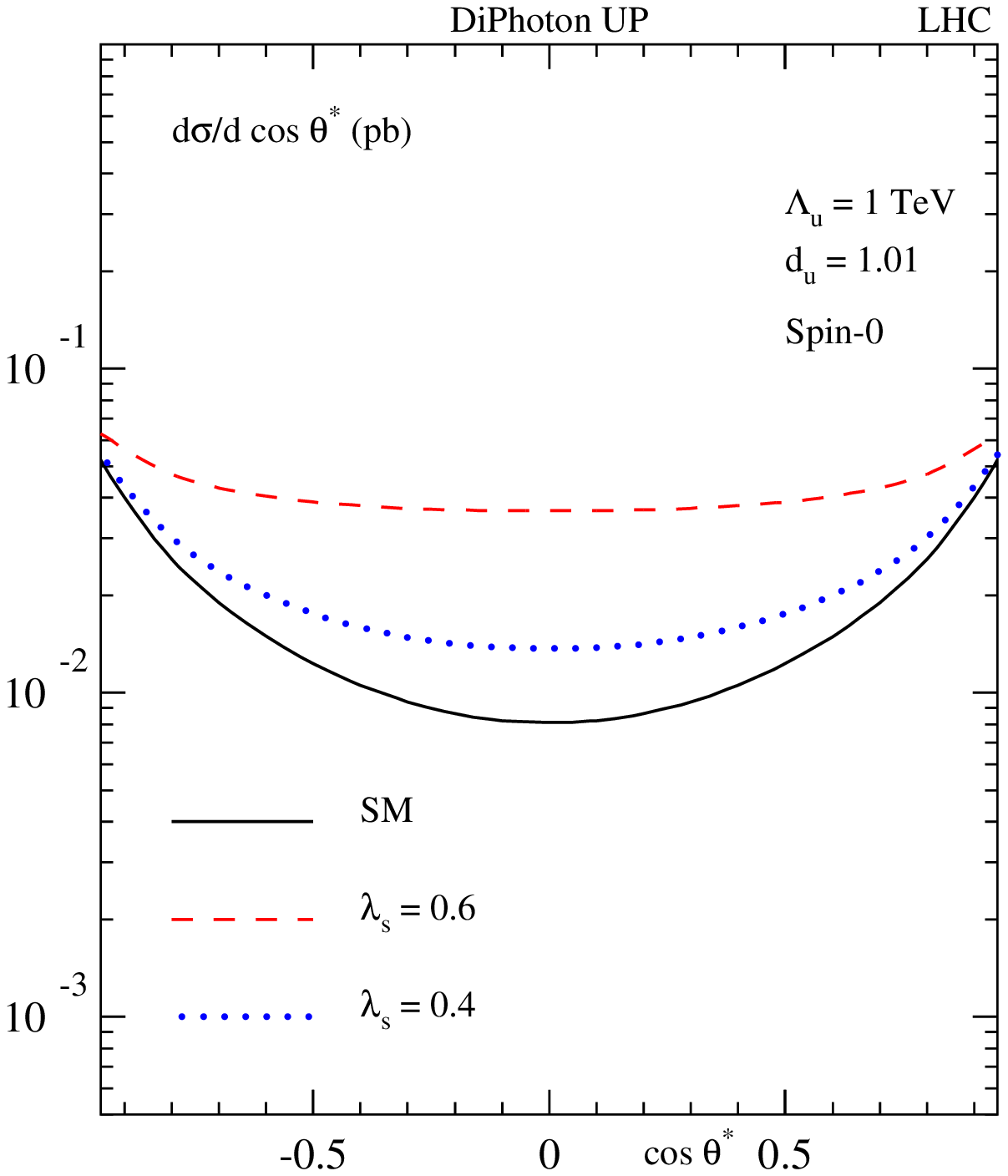,width=8cm,height=9cm,angle=0}
\epsfig{file=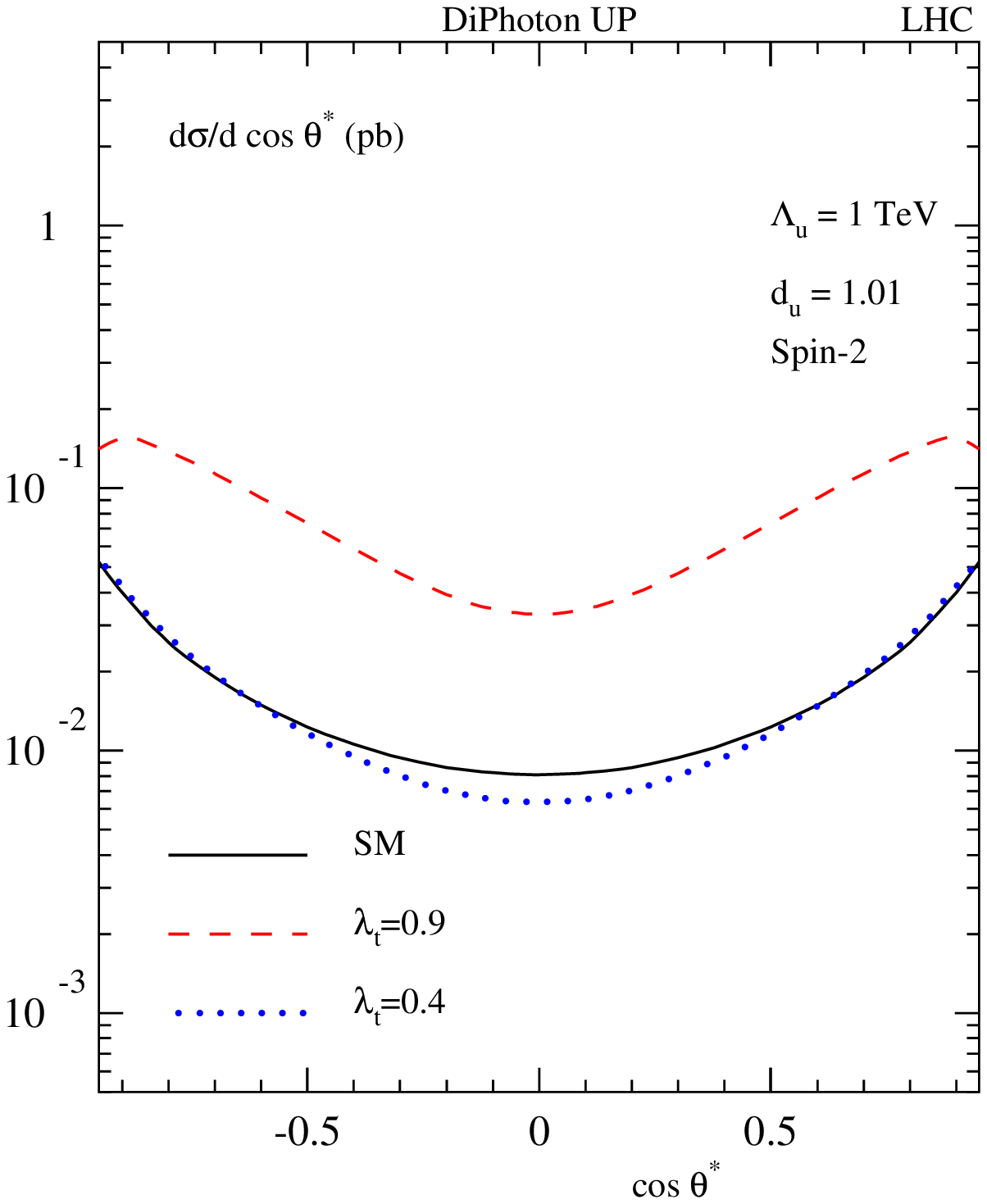,width=8cm,height=9cm,angle=0}}
\caption{ {Angular distributions $d\sigma/d{\cos}~\theta^*$ of the photons
for spin-0 (left) and spin-2 (right) with $\Lambda_{\cal U}=1$ TeV and
$d_{\cal U}=1.01$.  We have taken couplings $\lambda_s=0.6, 0.4$ and
and $\lambda_t=0.9, 0.4$ integrating Q in the range
600 GeV $ < Q < 0.9 \Lambda_{\uu}$.}}
\label{angulardistribution}
\end{figure}

The angular distributions for both spin-2 and spin-0 are 
given in Fig.~\ref{angulardistribution}.  These distributions for spin-2 
and spin-0 differ both in magnitude and in structure. For spin-2 case 
with $\lambda_t=0.9$, we find that  
the unparticle effects show up significantly. 
While for $\lambda_t=0.4 $, the negative interference term 
dominates over the pure unparticle contribution bringing down the distributions.
On the other hand, the scalar unparticle contribution 
for this choice of parameters is significant for both $\lambda_s = 0.4,0.6$.

\subsection{Rapidity}
       In this section, we consider the rapidity distributions of the
di-photon system ($d\sigma/dY$) as well as of individual final state photons 
($d\sigma/dy^{\gamma}$).  Rapidity of the di-photon system is defined as
\begin{eqnarray}
Y = \frac{1}{2} \log \left( \frac{p_2 . q}{p_1 .q} \right)\,,
\end{eqnarray}
here $q= p_3 + p_4$.
For the rapidity distributions of the individual photons, we have to replace $q$ in the
above equation by their momenta.
\begin{figure}[htb]
\centerline{
\epsfig{file=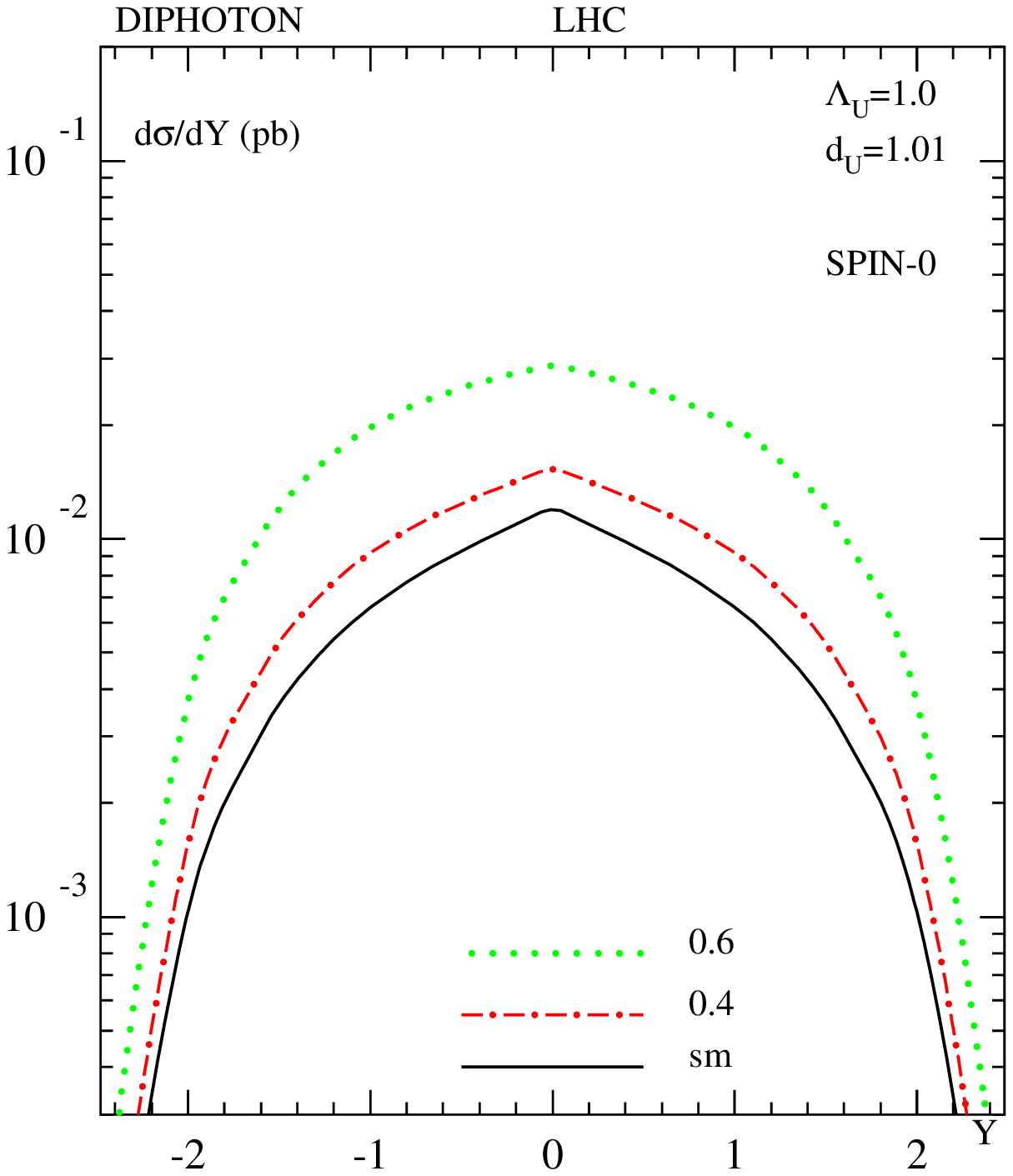,width=8cm,height=9cm,angle=0}
\epsfig{file=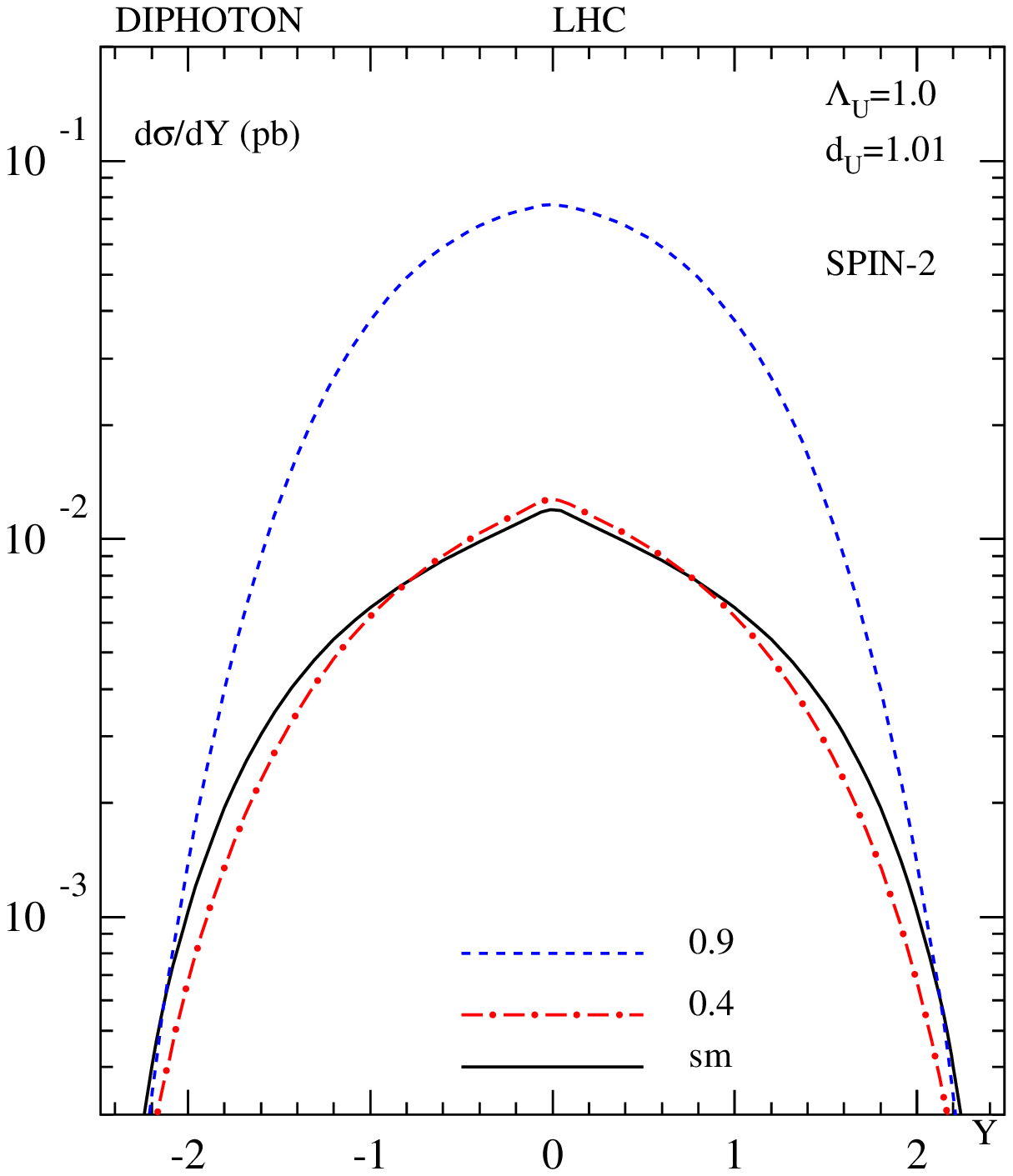,width=8cm,height=9cm,angle=0}}
\caption{ {Rapidity distributions $d\sigma/dY$ of the di-photon system
for spin-0 (left) and spin-2 (right) with $\Lambda_{\cal U}=1$ TeV and 
$d_{\cal U}=1.01$.  We have taken the couplings to be $\lambda_s=0.6, 0.4$ and
$\lambda_t=0.9, 0.4$ with Q in the region
600 GeV $< Q < 0.9 \Lambda_{\cal U}$.}}
\label{RapidityY}
\end{figure}

We have presented the rapidity distributions of di-photon system for both 
SM $+$ scalar unparticle and SM $+$ tensor unparticle in Fig.~\ref{RapidityY}.  
We have chosen $d_{\cal U}=1.01$ and $\Lambda_{\cal U}=1 $ TeV.  We find that the deviation 
from the SM is large in the central region (Y=0). 
For the spin-0 case and for both the choices $\lambda_s=0.6$ and $0.4$, the unparticle effect is
quiet large whereas for the spin-2 case, this is true only for 
larger $\lambda_t$.  

\begin{figure}[htb]
\centerline{
\epsfig{file=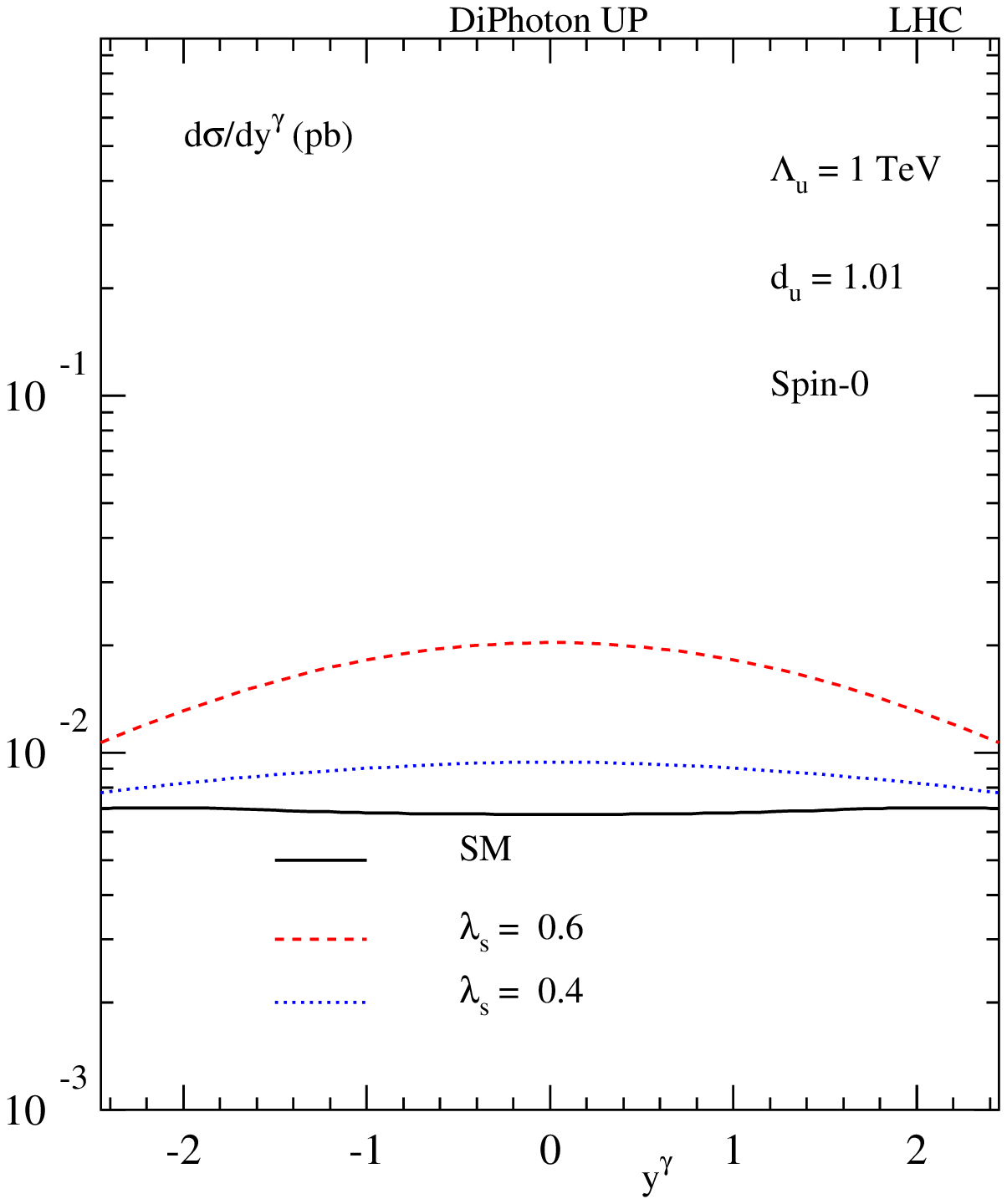,width=8cm,height=9cm,angle=0}
\epsfig{file=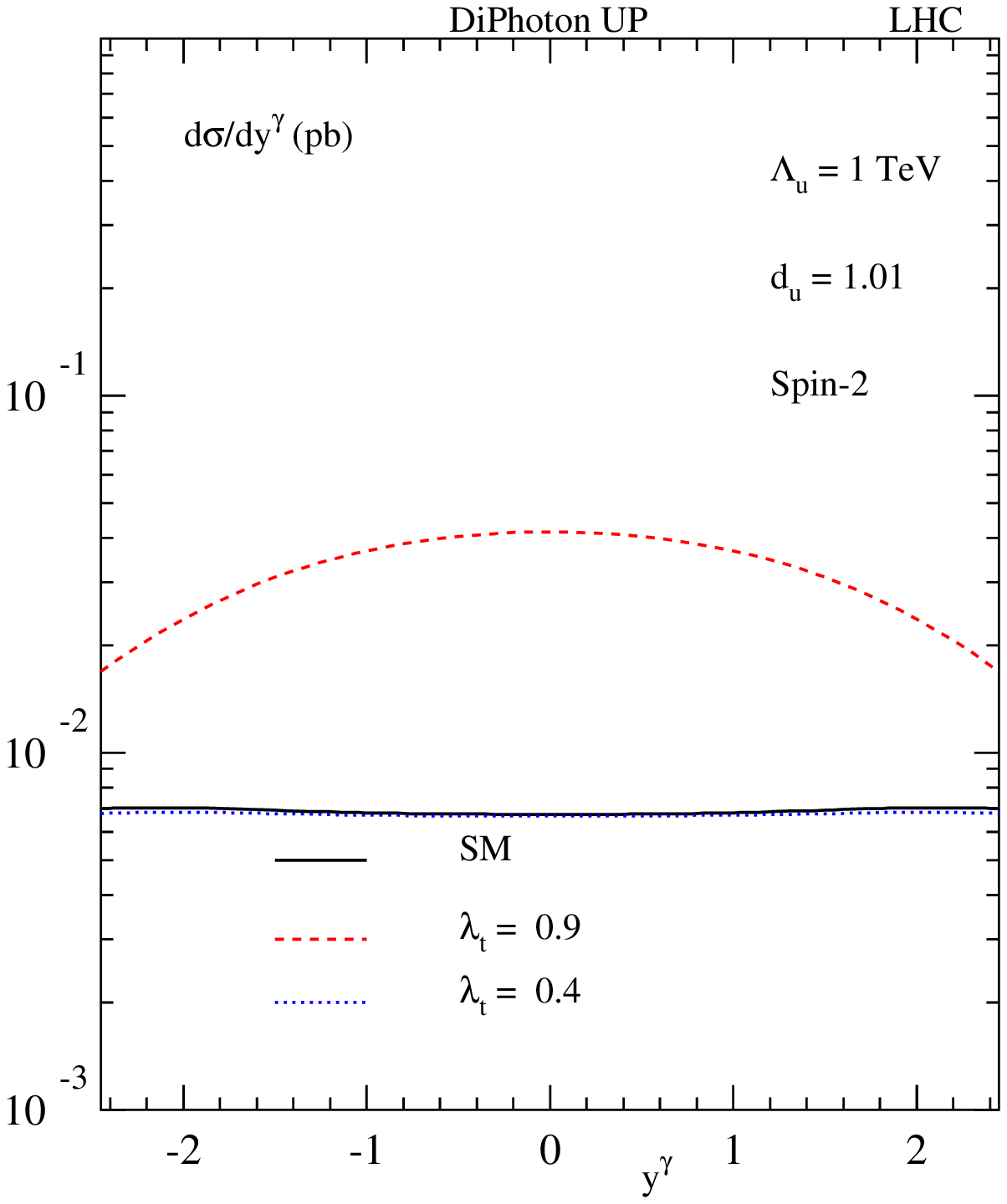,width=8cm,height=9cm,angle=0}}
\caption{ {Rapidity distributions $d\sigma/d y^{\gamma}$ of the photons
for spin-0 (left) and spin-2 (right) with $\Lambda_{\cal U}=1$ TeV and 
$d_{\cal U}=1.01$. We have taken the couplings to be $\lambda_s=0.6, 0.4$ and
$\lambda_t=0.9, 0.4$, with Q in the range
$600 { GeV } < Q < 0.9 \Lambda_{\cal U}$.}}
\label{rapidityyg}
\end{figure}
      We have presented the rapidity of the individual final state photons 
including the unparticle contributions 
in Fig.~\ref{rapidityyg} for $d_{\cal U}=1.01$ and $\Lambda_{\cal U}=1$ TeV.  
The scalar unparticle contribution with $\lambda_s=0.6$ is again large in the central region
whereas for the tensor 
unparticle case to see the larger effect  $\lambda_t$ has to be closer to $1$. 
For smaller values of $\lambda_{s,t}$, that is below $0.4$, the effects are unnoticeable.

\section{Conclusions}

In this paper we have studied the effects of scalar and tensor unparticles 
on the di-photon production at the LHC.  Various kinematical distributions are analysed 
to see their effects.
The spin-0 and spin-2 unparticle effects could be clearly distinguishable from SM 
background differing by around an order of magnitude in most of the 
distributions that we have considered.  
We have essentially three unknown parameters in this model,
namely the scale $\Lambda_{\cal U}$, the scaling dimension $d_{\cal U}$ and the coupling 
$\lambda_\kappa$.  
The impact of these parameters
on these distributions is studied in detail.
We find that the effects of scalar unparticle 
(Fig.~\ref{lambda}) with the coupling $\lambda_s$ closer to $1$ emerge even at 
lower energies.
We have chosen $\lambda_s$ below 0.6 for 
$d_{\cal U}=1.01$ and $\Lambda_{\cal U}=1$ TeV for our analysis. 
In case, the nature has both scalar and tensor unparticles, our results 
can give cross sections without any further work as these two do not
interfere.  We conclude by noting that the di-photon production can be used to unravel various
effects coming from this new model with unparticles.

\vspace{.5cm}
\noindent
{\bf Acknowledgments:}\\
M.C.Kumar (SRF) would like to thank CSIR, New Delhi, India for the financial
support. Anurag would like to thank Theory Division, SINP, Kolkata for 
the hospitality where a part of this work was done.
PM would like to thank the Abdus Salam ICTP for hospitality under the
Associates program, when this work was being completed.





\begin{thebibliography}{99}

\bibitem{bz}
  T.~Banks and A.~Zaks,
  Nucl.\ Phys.\  B {\bf 196}, (1982) 189.

\bibitem{hep-ph/0703260}
  H.~Georgi,
  Phys.\ Rev.\ Lett.\ {\bf 98}, (2007) 221601, arXiv:hep-ph/0703260.

\bibitem{hg2-kingman}
  H.~Georgi,
  Phys.\ Lett.\ B {\bf 650}, (2007) 275, arXiv:0704.2457 [hep-ph];
  K.~Cheung, W.~Y.~Keung and T.~C.~Yuan,
  arXiv:0704.2588 [hep-ph].


\bibitem{susy}
  P.~J.~Fox, A.~Rajaraman and Y.~Shirman,
  arXiv:0705.3092 [hep-ph];
%
H.~Zhang, C.~S.~Li, Z.~Li,
arXiv:0707.2132 [hep -ph];
%
Y.~Nakayama,
arXiv:0707.2451 [hep -ph];
%
N.~G.~Deshpande, X.~G.~He, J.~Jiang,
arXiv:0707.2959 [hep -ph];
%
T.~A.~Ryttov, F.~Sannino,
arXiv:0707.3166 [hep -ph].

\bibitem{hooman}
H.~Davoudiasl,
arXiv:0705.3636 [hep-ph];
S.~Hannestad, G.~Raffelt, Y.~Y.~Y.~Wong,
arXiv:0708.1404 [hep-ph],
  A.~Freitas and D.~Wyler,
  arXiv:0708.4339 [hep-ph].

\bibitem{flavor}
  C.~H.~Chen and C.~Q.~Geng,
  arXiv:0705.0689 [hep-ph];
%
  G.~J.~Ding and M.~L.~Yan,
  arXiv:0705.0794 [hep-ph];
  T.~M.~Aliev, A.~S.~Cornell and N.~Gaur,
  arXiv:0705.1326 [hep-ph];
%
  X.~Q.~Li and Z.~T.~Wei,
  \ Phys.\ Lett.\ B {\bf 651} (2007) 380-383 arXiv:0705.1821 [hep-ph];
%
  C.~D.~Lu, W.~Wang and Y.~M.~Wang,
  arXiv:0705.2909 [hep-ph];
%
  T.~M.~Aliev, A.~S.~Cornell and N.~Gaur,
   JHEP 0707:072,2007, arXiv:0705.4542 [hep-ph];
%
C.~H.~Chen, C.~Q.~Geng,
arXiv:0706.0850 [hep -ph];
%
        Roman Zwicky,
        arXiv: 0707.0677;
%
R.~Mohanta, A.~K.~Giri,
arXiv:0707.1234 [hep -ph];
%
C.~S.~Huang, X.~H.~Wu,
arXiv:0707.1268 [hep -ph];
%
A.~Lenz,
arXiv:0707.1535 [hep -ph];
%
  R.~Mohanta, A.K.Giri,
  arXiv:0707.3308 [hep-ph].
  C.~H.~Chen and C.~Q.~Geng,
  arXiv:0709.0235 [hep-ph].


 

\bibitem{neutrino}
S.~Zhou
arXiv:0706.0302 [hep -ph];
X.~Q.~Li, Y.~Li, Z.~T.~Wei,
arXiv:0707.2285 [hep -ph];
D.~Majumdar,
arXiv:0708.3485 [hep-ph],
  L.~Anchordoqui and H.~Goldberg,
  arXiv:0709.0678 [hep-ph].


\bibitem{Stephanov}
  M.~A.~Stephanov,
  arXiv:0705.3049 [hep-ph].

\bibitem{nath}
H.~Goldberg, P.~Nath,
arXiv:0706.3898 [hep-ph]

\bibitem{kikuchi}
T.~Kikuchi, N.~Okada,
arXiv:0707.0893 [hep -ph].

\bibitem{cheung}
K.~Cheung, W.~Keung, T.~Yuan,
arXiv:0706.3155 [hep -ph]

\bibitem{PM}
Prakash Mathews, V.~Ravindran,
Phys.Lett.{\bf B657}:198-206,2007,
arXiv:0705.4599 [hep-ph]

\bibitem{others}
%
M.~Luo and G.~Zhu,
arXiv:0704.3532 [hep-ph];
Y.~Liao,
arXiv:0705.0837 [hep-ph];
%
N.~Greiner, 
arXiv:0705.3518 [hep-ph];
%
D.~Choudhury, D.~K.~Ghosh, Mamta,
arXiv:0705.3637 [hep-ph];
%
Shao-Long Chen, Xiao-Gang He,
arXiv:0705.3946 [hep-ph];
%
G.~J.~Ding, M.~L.~Yan,
arXiv:0706.0325 [hep -ph];
%
Y.~Liao, J.~Y.~Liu,
arXiv:0706.1284 [hep -ph];
%
M.~Bander, J.~L.~Feng, A.~Rajaraman, Y.~Shriman,
arXiv:0706:2677 [hep-ph];
%
T.~G.~Rizzo,
arXiv: 0706.3025 [hep-ph];
S.~L.~Chen, X.~G.~He and H.~C.~Tsai,
arXiv:0707.0187 [hep -ph];
%
J.~J.~Bij, S.~Dilcher,
arXiv:0707.1817 [hep -ph];
%
D.~Choudhury, D.~K.~Ghosh,
arXiv:0707.2074 [hep -ph];
%
A.~Delgado, J.~R.~Espinosa, M.~Quiros,
arXiv:0707.4309 [hep-ph];
G.~Cacciapaglia, G.~Marandella, J.~Terning,
arXiv:0708.0005 [hep-ph];
%
M.~Neubert,
arXiv:0708.0036 [hep -ph];
%
M.~X.~Luo, W.~Wu, G.~H.~Zhu
arXiv:0708.0671 [hep -ph].
%
N.~G.~Deshpande, S.~D.~H.~Hsu, J.~Jiang
arXiv:0708.2735 [hep-ph];
%
G.~Bhattacharyya, D.~Choudhury, D.~K.~Ghosh,
arXiv: 0708.2835 [hep-ph];
%
Yi Liao,
arXiv:0708.3327 [hep-ph];
%
A.~T.~Alan, N.~K.~Pak,
arXiv:0708.3802;
%
  T.~i.~Hur, P.~Ko and X.~H.~Wu,
  arXiv:0709.0629 [hep-ph].
  I.~Gogoladze, N.~Okada and Q.~Shafi,
  arXiv:0708.4405 [hep-ph].
 

\bibitem{Binoth:1999qq}
  T.~Binoth, J.~P.~Guillet, E.~Pilon and M.~Werlen,
  Eur.\ Phys.\ J.\  C {\bf 16} (2000) 311
  [arXiv:hep-ph/9911340].

\bibitem{yuan}
    C.~Balazs, E.~L.~Berger, P.~M.~Nadolsky, C.~P.~Yuan
    Phys. Rev. D76:013009, 2007

\bibitem{eboli}
      O.~J.~P.~Eboli, T.~Han, M.~B.~Magro and P.~G.~Mercadante
      Phys. Rev. D61:094007, 2000;
%
      K.~Cheung, Greg L.~Landsberg
      Phys. Rev D62 (2000) 076003
\bibitem{ex}
CDF Collaboration,
PRL 95, 091801 (2005);
arXiv:0707.2294 [hep-ex]

\bibitem{bern}
Z.~ Bern, L.~ Dixon, C.~ Schmidt,
Phys.Rev. D66:074018, 2002; hep-ph/0206194

\bibitem{mack}
    G. Mack Commun. Math. Phys. 55, 1 

\bibitem{tdr}
    CMS Collaboration Physics Technical Design Report, 
    J. Phys. G: Nucl. Part. Phys. 34 995-1579.

\bibitem{Martin:2002dr}
  A.~D.~Martin, R.~G.~Roberts, W.~J.~Stirling and R.~S.~Thorne,
  Phys.\ Lett.\  B {\bf 531} (2002) 216
  [arXiv:hep-ph/0201127].

\end{thebibliography}
\end{document}